\documentclass[journal]{IEEEtran}
\usepackage{amsmath,amssymb,amsfonts}
\usepackage{xcolor}
\usepackage{multicol,multirow}
\usepackage{booktabs}
\usepackage{bigstrut}
\usepackage{rotating}
\usepackage{adjustbox}

\begin{document}
\title{LVS-Net: A Lightweight Vessels Segmentation Network for Retinal Image Analysis}
\author{Mehwish Mehmood, Shahzaib Iqbal \IEEEmembership{Member, IEEE} ,Tariq Mahmood Khan \IEEEmembership{Member, IEEE} ,Ivor Spence and Muhammad Fahim
\thanks{Mehwish Mehmood, Muhammad Fahim, and Ivor Spence are currently working with School of Electronics, Electrical Engineering and Computer Science, Queen's University Belfast, UK (e-mail:mmehmood01@qub.ac.uk, m.fahim@qub.ac.uk, I.spence@qub.ac.uk). }
\thanks{Shahzaib Iqbal is working with the Department of Computing, Abasyn University, Islamabad, Pakistan (e-mail: shahzeb.iqbal@abasynisb.edu.pk).}
\thanks{Tariq Mahmood Khan is with the School of Computer Science and Engineering, University of New South Wales, Sydney, Australia}
}
\maketitle

\begin{abstract}
The analysis of retinal images for the diagnosis of various diseases is one of the emerging areas of research. Recently, the research direction has been inclined towards investigating several changes in retinal blood vessels in subjects with many neurological disorders, including dementia. This research focuses on detecting diseases early by improving the performance of models for segmentation of retinal vessels with fewer parameters, which reduces computational costs and supports faster processing. This paper presents a novel lightweight encoder-decoder model that segments retinal vessels to improve the efficiency of disease detection. It incorporates multi-scale convolutional blocks in the encoder to accurately identify vessels of various sizes and thicknesses. The bottleneck of the model integrates the Focal Modulation Attention and Spatial Feature Refinement Blocks to refine and enhance essential features for efficient segmentation. The decoder upsamples features and integrates them with the corresponding feature in the encoder using skip connections and the spatial feature refinement block at every upsampling stage to enhance feature representation at various scales. The estimated computation complexity of our proposed model is around 29.60 GFLOP with 0.71 million parameters and 2.74 MB of memory size, and it is evaluated using public datasets, that is, DRIVE, CHASE\_DB, and STARE. It outperforms existing models with dice scores of 86.44\%, 84.22\%, and 87.88\%, respectively.

\end{abstract}

\section{Introduction} \label{introduction}

\IEEEPARstart{T}{he} retina, a morphological and functional extension of the brain, is part of the neurological system; some of its cells directly connect to the brain \cite{soomro2016automatic, khan2016automatic, khan2019boosting, soomro2018impact, khan2019generalized}. Retinal blood vessels are the only vascular network in the human body that is directly and non-invasively observable \cite{khan2019ggm, khawaja2019improved, khawaja2019multi, khan2019use, khan2020shallow}. Retinal vascular structural abnormalities are significant for their clinical implications in identifying systemic and neurological disorders, including dementia, such as Alzheimer's disease \cite{khan2020exploiting, khan2020semantically, naveed2021towards, abdullah2021review, imtiaz2021screening}. This has led to a growing interest in investigating how early Alzheimer's disease causes retinal changes, as retinal vessel health may reflect the cerebral vasculature's condition due to the homology between the retina and the cerebral microvasculature \cite{wiseman2023retinal}. An estimated 16 million people worldwide are affected by retinal vascular conditions, with Alzheimer's being the second most common retinal vascular condition after diabetic retinopathy \cite{mirzaei2020alzheimer, khan2021residual, khan2022width,naqvi2023glan,khan2023retinal}.

Medical image segmentation plays a crucial role in healthcare by enabling precise delineation of anatomical structures and pathological regions, which aids in disease diagnosis, treatment planning, and monitoring \cite{khan2024lmbf, iqbal2024tesl,iqbal2024euis,iqbal2025tbconvl, naveed2024ad, farooq2024lssf, muhammad2024advancing}. Fundus imaging offers a non-invasive and efficient means of studying the cerebral microvasculature and its relationship to dementia. Retinal images are easy to acquire and growing evidence suggests that this microvascular network may represent the cerebral microvasculature \cite{burns2021imaging,iqbal2023robust,manan2023semantic}. Retinal image analysis has been dramatically improved by the development of image processing and machine learning technologies, which have improved disease diagnosis and monitoring \cite{Mehmood_2024_CVPR,khan2023feature,abbasi2023lmbis,iqbal2023ldmres,mehmood2019vessel,matloob2024lmbis,javed2024region}. The retinal vascular segmentation of these images is essential for monitoring anatomical changes and possibly diagnosing diseases like Alzheimer's. Retinal image analysis is tedious as their color or gray level varies from one part to another due to the morphology of the retinal structure and different features, resulting in inaccurate output.

Retinal vascular segmentation remains challenging despite advancements, particularly in terms of vessel distinction \cite{iqbal2022recent,khan2021rc,khan2022leveraging,khan2022t,iqbal2022g,arsalan2022prompt,khan2022mkis}.  This research introduces LVS-Net, a novel lightweight encoder-decoder model that improves the effectiveness of disease screening by segmenting retinal vessels. Our proposed lightweight model integrates the Focal Modulation Attention Module ($\texttt{FMAM}$) \cite{yang2022focal} and the spatial feature refinement block ($\texttt{SFRB}$) \cite{wang2019spatial} into the bottleneck to refine and improve essential features for efficient segmentation. The decoder upsamples features and integrates them with the corresponding feature in the encoder using skip connections and $\texttt{SFRB}$ to enhance feature representation at various scales and to fine-tune the segmented image's overall representation. The suggested model is evaluated on public datasets, that is, DRIVE \cite{qureshi2013manually}, CHASE\_DB \cite{7530915}, and STARE \cite{STAREDataset}. The results show a notable improvement in retinal vascular segmentation over existing accuracy, efficiency, and complexity methods.

The primary contributions of this paper include:

\begin{itemize} 

\item Introducing LAV-Net, a novel lightweight encoder-decoder model specifically designed for segmentation of retinal vessels. LAV-Net is highly effective and performs exceptionally well with a relatively small number of learnable parameters (only 0.71M).

\item Implementing multi-class segmentation to identify arteries and veins from retinal images, facilitating the model to retrieve multiple features simultaneously while improving efficiency.

\item Integrating focal modulation attention and spatial feature refinement blocks at the encoder-decoder bottleneck to refine important features guarantees that more detailed information is transferred for efficient segmentation. Crucial details are maintained and improved during the decoding process by introducing $\texttt{SFRB}$ in the decoder block that combines robust upsampling techniques with feature aggregation. \end{itemize}

The paper is organized as follows: Section \ref{Related Work} contains the work related to retinal vessel segmentation. The methodology of the suggested model is described in Section \ref{Methodology}. The experiments and findings are covered in Section \ref{Results}. Lastly, Section \ref {Conclusion} contains the conclusion.

\section{Related Work}\label{Related Work}
In the past few years, deep learning-based techniques are common for segmenting retinal blood vessels and have outstanding accuracy that suppressed traditional segmentation methods.
The recent advancement in this area is discussed below.

\subsection{Retinal Vessel Segmentation}
Chowdhury {\it et al.} \cite{chowdhury2022msganet} developed MSGANet-RAV, a U-shaped encoder-decoder network that segments and classifies retinal vessels. Although it performed well, particularly in handling vessel crossings, it still demonstrated limitations in accurately segmenting small vessels and other detailed structures. Hemelings {\it et al.} \cite{hemelings2019artery} utilized a U-Net architecture, achieving high accuracy, but leaving room for improvement in the discrimination of small vessels. Lyu {\it et al.}\cite{Lyu2022} presented a convolutional neural network (CNN) model comparable to U-Net for the segmentation of binary vessels and the assessment of fractal dimensions. Xu {\it et al.} \cite{XU2022695} proposed a dual channel asymmetric CNN based on the U-Net model based on pre-processing scale and orientation features. The combination of the segmentation results from both channels resulted in comprehensive and complementary information. However, the effectiveness of these techniques is limited by the pathological variability present in clinical images and the various scales of vascular geometry. Additionally, these variations may incorporate supplementary structural elements, leading to a higher number of network learnable parameters and, as a result, increased GPU memory consumption. 

Unlike CNN-based architectures, transformers have recently been adopted in many computer vision tasks. Vision transformers (ViTs) \cite{dosovitskiy2020image} have received a substantial research interest, and various subsequent approaches have been presented that expand on ViTs. They adjusted the architecture by cascading numerous transformer layers instead of CNN-based architectures. Due to the powerful representation learned from pre-trained backbones, numerous algorithms for semantic segmentation that incorporate ViT backbones demonstrate impressive results \cite{zhang2022segvit, zheng2021rethinking, strudel2021segmenter, ranftl2021vision}. Although ViTs yield robust outcomes, they are computationally expensive.

Using lightweight CNNs is one potential way to overcome these problems. There are several advantages to using lightweight techniques in medical imaging, including faster processing speeds, lower memory requirements, better portability, lower computational costs, and lower power usage. These advantages make them an attractive option for several applications, which has recently attracted the interest of most academics. Wentao {\it et al.} \cite{liu2022full} introduce FR-UNet, a novel segmentation technique designed to improve segmentation accuracy and vessel connectivity. A dual-threshold iterative approach is used to capture weak vessel pixels, and the network retains full image resolution while enhancing feature extraction through a multiresolution convolution mechanism. FR-UNet tends to create false positives during the segmentation of thin vessels. The Dense-Inception U-Net was proposed by Zhang {\it et al.}\cite{zhang2020dense}, and utilizes a tiny encoder along with a lightweight backbone and dense module to capture high-level semantic information. 

Several researchers have developed lightweight networks specifically for the segmentation of medical images. However, achieving minimal model complexity and rapid inference while maintaining outstanding performance remains challenging in medical imaging. NnU-Net \cite{isensee2021nnu} preprocesses the data and postprocesses the segmentation outcomes to boost network flexibility; however, this method increases the model parameters. The lightweight V-Net \cite{lei2020lightweight} allows for efficient segmentation and uses a lower number of parameters utilizing point-wise and depth-wise convolution; nevertheless, it fails to accelerate the inference process of the model. Furthermore, using multimodal magnetic resonance imaging (MRI), Tarasiewicz {\it et al.}\cite{tarasiewicz2020lightweight} trained multiple tiny networks across all image channels to create lightweight U-Nets to precisely identify brain tumors. By replacing all convolutional layers in conventional U-Net with pyramidal convolution, PyConvU-Net \cite{li2021pyconvu} increases segmentation accuracy and requires a smaller number of parameters. Nevertheless, PyConvU-Net's inference time remains inadequate. CNN architectures that are lightweight and effective for segmenting retinal blood vessels are G-Net Light \cite{iqbal2022g}, PLVS-Net \cite{arsalan2022prompt}, TBConvL-Net\cite{iqbal2025tbconvl}, Lmbf-net \cite{khan2024lmbf} and MKIS-Net \cite{khan2022mkis}. There are two significant drawbacks to existing lightweight methods. Firstly, they do not match the state-of-the-art methods in terms of performance, and secondly, they cannot generalize. Existing methods, including lightweight methods, unable to detect multi-retinal features with state-of-the-art results \cite{howard2017mobilenets}.

\subsection{Retinal Arteries Veins Segmentation}
Morano {\it et al.} \cite{morano2021simultaneous} presented a novel approach that uses fully convolutional neural networks (FCNN) and a unique loss function of `Binary cross entropy by 3' (BCE3) for simultaneous segmentation and classification of retinal arteries and veins. This approach performs exceptionally well in handling vessel crossings, but still has room for improvement. Shi {\it et al.}\cite{shi2024one} present a novel one-shot method to segment the retinal arteries and veins that uses fundus fluorescein angiography (FFA) and color fundus photography (CFP) with cross-modal pre-training. This method trained a GAN to produce soft segmentation AV using CFP inputs. There is still a need for improvement, as the approach indicated a possible loss of small vessel information, even with its capacity to manage insufficient data. LUNet, a unique deep learning architecture for segmentation of the arteries and veins in high-resolution fundus images, was introduced by J. Fhima {\it et al.} \cite{fhima2024lunet}. In order to improve the receptive field, the model has a particular double-dilated convolutional block. LUNet relies on high-quality images, which can affect segmentation accuracy in cases where image quality is poor. Despite good performance on high-resolution datasets, challenges in processing large-scale datasets and small vessels persisted. \\

Danli {\it et al.} presented the Retina-based microvascular health assessment system (RMHAS) using a multibranch U-Net to segment the optic disc, veins, and arteries. Real-time application may be limited by the multi-step RMHAS process, which includes image quality evaluation and multi-branch segmentation. This procedure may also increase computational complexity and processing time. Jingfei Hu {\it et al.} \cite{hu2022multi} presented a novel multi-scale interactive network with an artery-venous (A/V) discriminator for classifying retinal arteries and veins. Using a special A/V discriminator, the network integrates multi-scale data and tackles common problems such as arteriovenous confusion and vascular discontinuity. Real-time implementation may be constrained by the method's complexity in integrating multi-scale data and implementing an A/V discriminator, which could raise computing costs.  
Wenao Ma {\it et al.} \cite{ma2019multi} presented a network in which the input module integrates domain knowledge from widely utilized vessel enhancement and retinal preprocessing methods. Specifically created for the network output block, a spatial activation mechanism uses vessel segmentation to improve A/V classification performance. Deep supervision is also incorporated into the network to help lower layers extract valuable data. Pre-processing is a significant method component which could limit generalization to different datasets and increase model complexity.\\

Segmentation of retinal vessels has advanced significantly. However, there are still many challenges with the existing models. Real-time implementation is challenging due to the high memory requirements and the longer inference time of complex models, which are computationally intensive \cite{li2020lightweight}. Although lightweight models have benefits such as reduced memory utilization and faster processing, they typically fail to incorporate complex processes like attention modules. This leads to a substantial gap in the development of lightweight models that incorporate attention processes for improved segmentation accuracy, particularly in the segmentation of retinal vessels \cite{chen2021retinal}.

\begin{figure*}[h!]
  \centering
    \includegraphics[width=\textwidth]{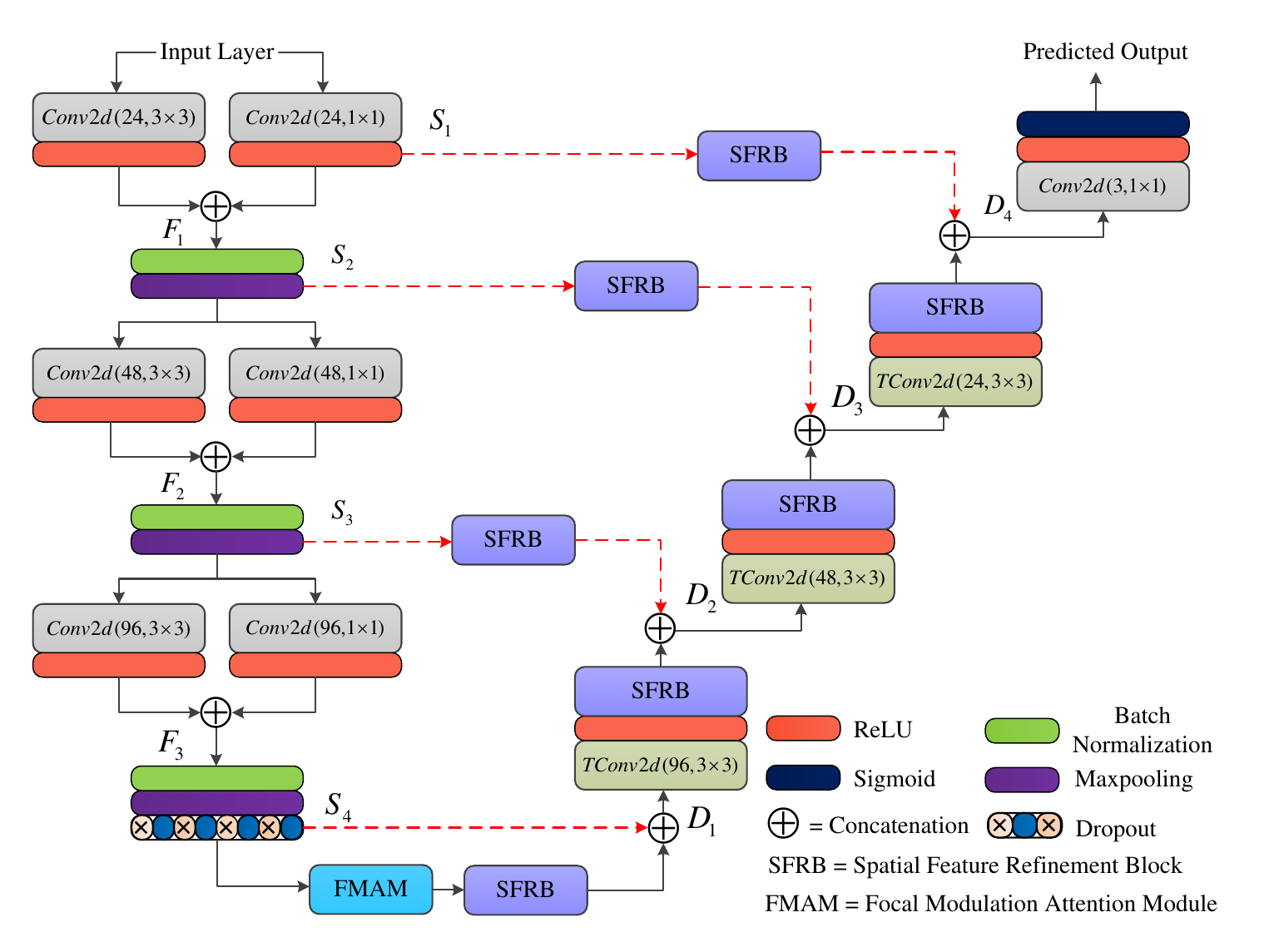} 
    \caption{Architecture of the proposed LVS-Net: begins with convolutional operations followed by a decoding path using transposed convolutions. Key elements include Focal Modulation Attention and Spatial Feature Refinement block in various stages of model for precise segmentation.}\label{LAV-Net}
\end{figure*}


\begin{figure*}
    \centering
\begin{tabular}{cc}
     \includegraphics[width=0.5\textwidth]{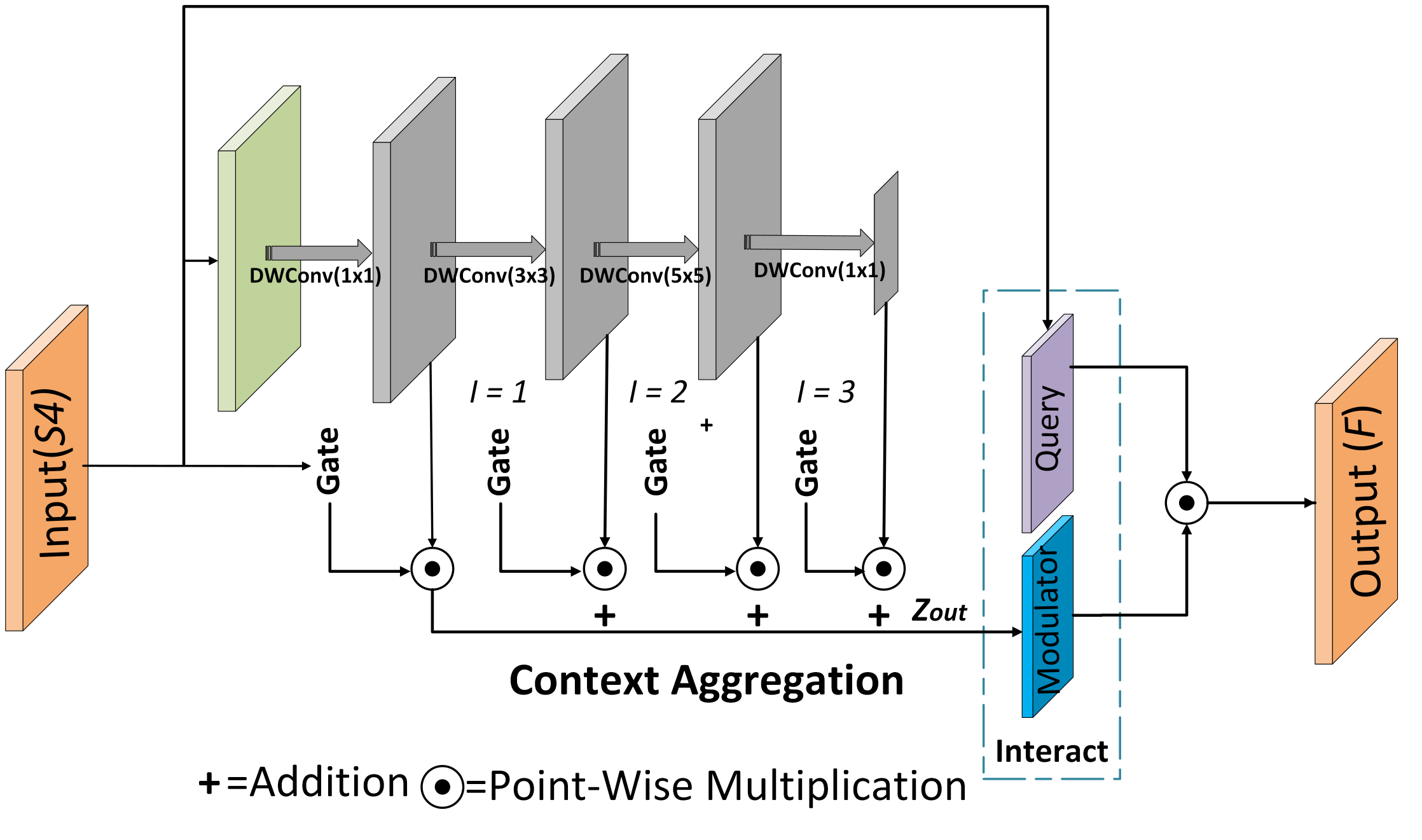} &  \includegraphics[width=0.5\textwidth]{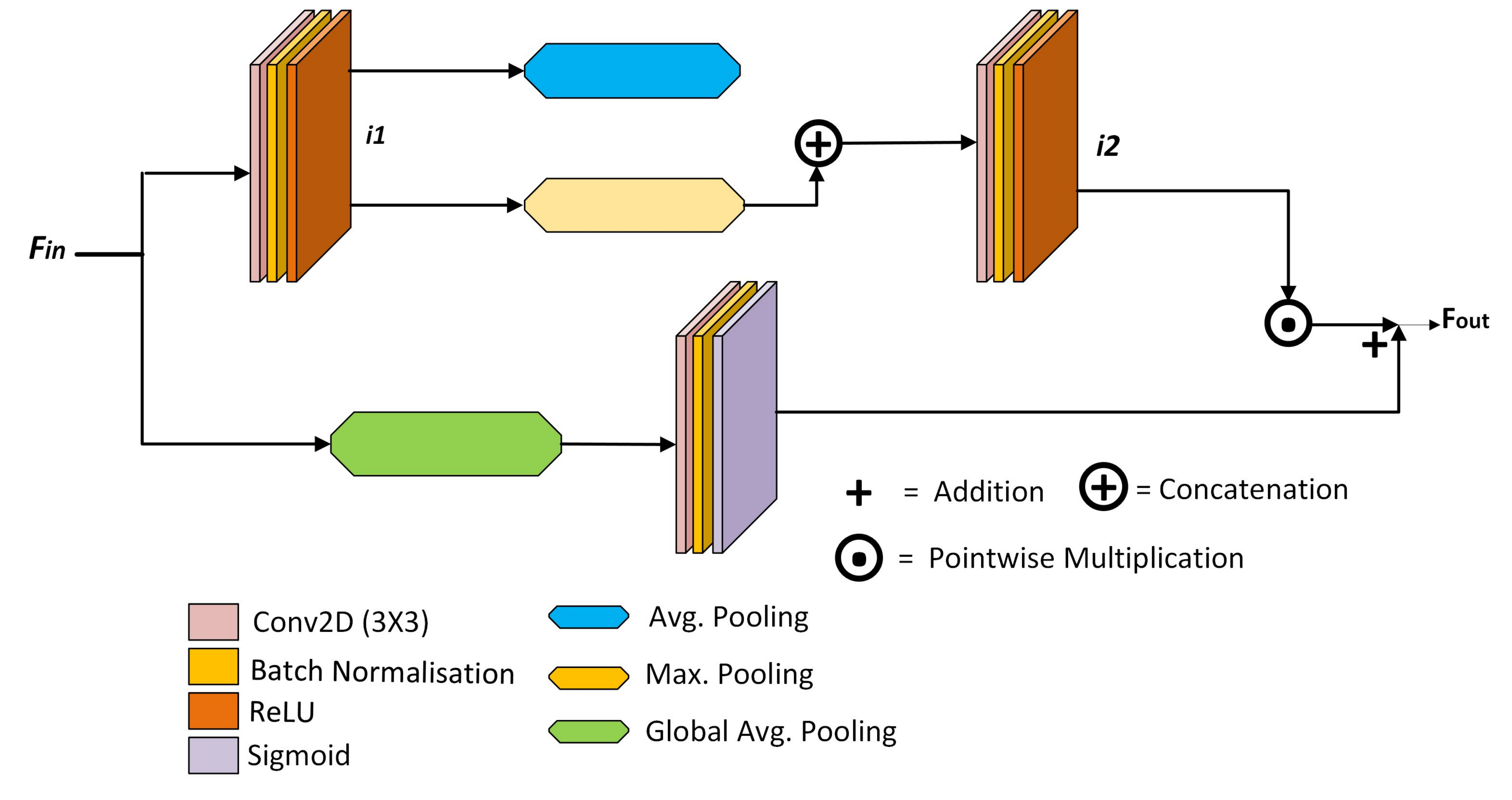}\\
     (a) & (b) \\
     
\end{tabular}
    \caption{Schematics of the proposed blocks: (a) Focal modulation attention module with context aggregation, (b) Spatial feature refinement block.}
    \label{FMAB}
\end{figure*}

\section{Methodology}\label{Methodology}
Our lightweight model is designed for the segmentation of retinal vessels with an encoder-decoder architecture. The images in the datasets used for model evaluation are limited in number. To address this issue, data augmentation is employed, which consists of rotating the images by 20 degrees and adjusting the contrast of the rotated images. The architecture of our proposed model is discussed in the following section. 

\subsection{Model Architecture}\label{subsec:atchitecture}
We present a lightweight encoder-decoder architecture, LVS-Net, to extract the retinal vessels from fundus images. The overall network architecture is depicted in Fig. \ref{LAV-Net}, which introduces multi-scale analysis and feature refinement inside the model. To enhance channel mapping and fine-tune the details, this network's bottleneck layer uses focal modulation along with spatial feature enhancement. The proposed LVS-Net comprises skip connections to preserve the original information from the encoder and use it during the decoding process. The output of the $1^{st}$ skip connection ($S_{1}$) is obtained by applying an activation function and a convolution operation on the RGB input image with $512\times512$ resolution, as given in Eq. \ref{eq:M1}.

\begin{equation}
   S_{1}= \mathit{ReLU}(C^{1\times 1,24}(Img_{512\times 512}))
    \label{eq:M1}
\end{equation}

Here, the operation $C^{1\times 1,24}$ denotes a convolution with $1\times 1$ kernel size and 24 output channels. $\mathit{ReLU}$ represents the activated Rectified Linear Unit (ReLU). The $1^{st}$ convolution block employs an activation function after a $3\times3$ convolution operation on $Img_{512\times 512}$. The resulting output is concatenated with ($S_{1}$), as shown in Eq. \ref{eq:M1a}. 
\begin{equation}
   F_{1}= \mathit{ReLU} (C^{3\times 3,24}(Img_{512\times 512}))\oplus  S_{1}
    \label{eq:M1a}
\end{equation}
Where $\oplus$ is the concatenation operation. The $2^{nd}$ skip connection employs a batch normalization on $F_{1}$ followed by a max-pooling layer to reduce the spatial dimensions, as shown in Eq. \ref{eq:M2}.

\begin{equation}
   S_{2}=\text{MaxPool}_{2 \times 2}\left ( \mathit{BN} \left (F_{1} \right)\right )
    \label{eq:M2}
\end{equation}

Where $\mathit{BN}$ is the batch normalization. The $2^{nd}$ convolution block and the $3^{rd}$ skip connection are mathematically represented  in Eq. \ref{eq:M3a} and Eq. \ref{eq:M3}, respectively.
\begin{equation}
   F_{2}= Re (C^{1\times 1,48}(S_{2}))\oplus  \mathit{ReLU} (C^{3\times 3, 48}(S_{2}))
    \label{eq:M3a}
\end{equation}

\begin{equation}
   S_{3}=\text{MaxPool}_{2 \times 2}\left ( \mathit{BN} \left ( F_{2}  \right )\right )
    \label{eq:M3}
\end{equation}
Finally, the $3^{rd}$ convolution block and $4^{th}$ skip connection are presented in Fig. \ref{eq:M4a} and Eq. \ref{eq:M4}, respectively.

\begin{equation}
  F_{3}=\mathit{ReLU} (C^{1\times 1,96}(S_{3}))\oplus \mathit{ReLU} (C^{3\times 3,96}(S_{3}))
    \label{eq:M4a}
\end{equation}

\begin{equation}
  S_{4}=Dr^{(0.5)}\left ( \text{MaxPool}_{2 \times 2}\left ( \mathit{BN} \left (F_{3} \right )\right ) \right )
    \label{eq:M4}
\end{equation}

Where $D^{(0.5)}$ is the dropout operation with a probability of 0.5. The proposed LVS-Net applies the $\texttt{FMAM}$ to refine the encoded features and improve the channel mapping after using max-pooling layers. Once the encoder features are refined by $\texttt{FMAM}$, they are processed through $\texttt{SFRB}$ and then concatenated with the actual information using skip connection to compute the output of the decoder block $1^{st}$ as shown in Eq. \ref{eq:M5}.

\begin{equation}
D_{1}=\mathcal{G} \left ( \mathcal{F}\left (S_{4} \right ) \right )\oplus S_{4}
    \label{eq:M5}
\end{equation}

Where, $\texttt{SFRB}$ and $\texttt{FMAM}$ are denoted as $\mathcal{G}$ and $\mathcal{F}$, respectively. \\

A transposed convolution operation is applied to $D_{1}$ to up-sample feature maps, followed by a ReLU activation and $\texttt{SFRB}$ to enhance spatial features. Subsequently, the feature maps are concatenated with the output of $\texttt{SFRB}$ applied to $S_{3}$ to preserve the original information as shown in Eq. \ref{eq:M6}.

\begin{equation}
D_{2}=\mathcal{G} \left ( \mathit{ReLU}\left ( T^{3\times 3,96}(D_{1}) \right ) \right )\oplus   \mathcal{G} \left (S_{3} \right )
    \label{eq:M6}
\end{equation}
Where the operation $T^{3\times 3,96}$ denotes a transposed convolution with $3\times 3$ kernel size and 96 output channels. The output of the $3^{rd}$ and $4^{th}$ decoder blocks ($D_{3}$ and $D_{4}$) are calculated as Eqs. \ref{eq:M7} - \ref{eq:M8}

\begin{equation}
D_{3}=\mathcal{G} \left ( \mathit{ReLU}\left ( T^{3\times 3,48}(D_{2}) \right ) \right )\oplus   \mathcal{G} \left (S_{2} \right )
    \label{eq:M7}
\end{equation}

\begin{equation}
D_{4}=\mathcal{G} \left ( \mathit{ReLU}\left ( T^{3\times 3,24}(D_{3}) \right ) \right )\oplus   \mathcal{G}\left (S_{1} \right )
    \label{eq:M8}
\end{equation}
The final predicted mask is computed by applying $1\times 1$ convolution operation, followed by a ReLU activation and sigmoid function, as shown in Eq. \ref{eq:M9}.

\begin{equation}
I_{out}=\sigma \left ( \mathit{ReLU} \left ( C^{1\times 1,1} \left ( D_{4} \right )\right ) \right )
    \label{eq:M9}
\end{equation}
 Where $\sigma$ is the sigmoid operation. To convert the predicted map from the decoder to a segmentation mask, F1-thresholding is used that maximizes the dice score. Furthermore, the dice loss function and Adam optimizer are used to train our model. Dice loss measures what percentage of overlap occurs between the segmented image \( S \) and the ground truth (GT) image \( G \):
 
\begin{equation}
\mathcal{L}_d(S, G) = 1 - \sum_{k=1}^{c} w_k \frac{2 \sum_{j=1}^{n} S(k, j) \cdot G(k, j)}{\sum_{j=1}^{n} S(k, j)^2 + \sum_{j=1}^{n} G(k, j)^2 + \xi}
\label{Loss}
\end{equation}

where \( w_k \) represents the weight of the \( k \) -th class, \( c \), \( n \) and \( \xi \) are the number of pixels, the number of classes, and a smoothing constant, respectively.

\subsection{Focal Modulation Attention Module}\label{subsec:FMAM}

To further improve the obtained feature information, $\texttt{FMAM}$ has been utilized between the encoder and the decoder. It is made up of three different elements, as Fig. \ref{FMAB}(a) illustrates. It initially employs input from the encoder to encode the visual details at short and long ranges using a sequence of depth-wise convolutional layers. Each layer in the stack extracts various levels of information, including local and global details. 
\begin{equation}
z^{\mathit(l)} = (\text{DWConv}(z^{\mathit(l)-1}))
\end{equation}
Where \(z^{\mathit(l)}\) is the output feature map at level \(\mathit(l)\) and \(\text{DWConv}\) is depth-wise convolution at level \(\mathit(l)\).\\

The global context is obtained as:
\begin{equation}
z^{L+1} = \text{AvgPool}(z^L)
\end{equation}

It refines input features through hierarchical context aggregation and modulation.The aggregated context features across all levels are combined as:
\begin{equation}
Z_\text{out} = \sum_{\mathit(l)=1}^{L+1} G^{\mathit(l)} \odot z^{\mathit(l)}
\end{equation}
where, \(G^{\mathit(l)}\) is gating weights at level \(\mathit(l)\), and \(\odot\) is element-wise multiplication.\\

Using the context as guidelines, these extracted data are utilized later to collect context attributes for each query token in a selective manner. The significance of each contextual data in the final representation of the query token is determined by a gate with learnable attention weight. The modulated output for each query token \(F_i\) is computed as:
\begin{equation}
F_i = q(S4_i) \odot h(Z_\text{out}),
\end{equation}
where i represents the spatial index in the feature map, \(q(S4_i)\) is query projection function and \(h(Z_\text{out})\) is modulator projection function. An element-wise multiplication is performed to fuse these combined context characteristics into the query token. A learnable weight matrix that is updated throughout the training defines this transformation. 

\subsection{Spatial Feature Refinement Block}\label{subsec:SFRB}
Pooling operations are used to reduce model complexity and computational overhead. Max-pooling retains the dominant features, while average-pooling retains low-frequency features for global context. Parallel paths of the max-pooling and average-pooling operations are implemented in $\texttt{SFRB}$ to integrate local and global features effectively.\\

In this approach, in Fig. \ref{FMAB}(b), the input feature map $\mathit{Fin}$ undergoes initial processing through convolutional layers, followed by batch normalization and ReLU activation represented in Eq. \ref{eq:SEM1}. 

\begin{equation}
    i_{1}=\mathit{ReLU}(\mathit{BN}(C^{3\times 3}(F_{in})))
    \label{eq:SEM1}
\end{equation}
Subsequently, the results of the max-pooling $ P^{3\times 3}_{\texttt{Max.}}$ and average-pooling $P^{3\times 3}_{\texttt{Avg.}}$ are concatenated, followed by convolution ($C$), batch normalisation ($\mathit{BN}$) and ReLU activation ($\mathit{ReLU}$) calculated by Eq. \ref{eq:SEM2}.

\begin{equation}
    i_{2}=\mathit{ReLU}(\mathit{BN}(C^{3\times 3}[P^{3\times 3}_{\texttt{Avg.}}(i_{1}) \oplus P^{3\times 3}_{\texttt{Max.}}(i_{1})]))
    \label{eq:SEM2}
\end{equation}


Furthermore, an additional pathway is introduced that incorporates global average pooling ($P^{\texttt{Global}}_{\texttt{Avg.}}$), convolutional operations, batch normalization, and sigmoid activation function ($\Sigma$) to generate attention coefficients to weight the results of parallel pooling. Subsequently, the weighted feature map is combined with the input to produce the output $F{out}$ of $\texttt{SFRB}$, as shown in Eq.\ref{eq:SEM3}.

\begin{equation}
    F_{out}=[\sigma(\mathit{BN}(C^{3\times 3}(P^{\texttt{Global}}_{\texttt{Avg.}}(F_{in})))) \odot i_{2}] + F_{in}
    \label{eq:SEM3}
\end{equation}

\section{Experiments and Results}\label{Results}

\subsection{Datasets}\label{subsec:datasets}
Our model was evaluated on DRIVE, STARE, and CHASE\_DB-DB datasets. There are forty colored retinal images in the DRIVE dataset \cite{qureshi2013manually}, each with a resolution of $565\times 584$ pixels (8 bits per channel). These images were taken with a Canon CR5 non-mydriatic 3CCD camera with a 45-degree viewing field. Each of the two subsets of the data set, the test set and the training set, consists of twenty images. Specifically, the test set benefits from two expert annotations, but the training set has access to one. STARE \cite{STAREDataset} dataset of retinal images widely employed in the development and evaluation of algorithms for the segmentation of retinal lesions. The Ophthalmic Image Analysis Laboratory created the University of California, Berkeley dataset. The STARE dataset includes retina scans of 13 different human eye diseases and a catalog of disease names and codes associated with each image. Manual expert annotations are available for the blood vessels and the optic nerve. It comprises 20 color fundus images with a $700 \times 605$ pixel resolution, acquired with a ``Topcon TRV-50'' camera. The images were obtained from patients with various retinal diseases and pre-processed to eliminate nonretinal areas. For each image, two independent expert graders provide ground-truth segmentation of retinal blood vessels. The CHASE\_DB dataset \cite{7530915} comprises 28 images, each with a resolution of $1024\times 1024$ pixels. Scanning laser ophthalmoscopy (SLO) produces high-resolution retinal images, making these images distinctive. Each image in the CHASE\_DB collection was captured using an EasyScan device (i-Optics Inc.) with a 45-degree field of view (FOV) that combines the SLO approach. A group of retinal image analysis professionals carefully annotated the vessels in this dataset.
\begin{table*}[h!]
\centering
\caption{Overview of datasets and their properties including the number of training and testing images, total and augmented images, original image resolution, field of view (FOV), and training details providing better insight into their application.}
 \resizebox{\textwidth}{!}{%
\begin{tabular}{llccccccl}
\toprule
\textbf{Feature} & \textbf{Dataset} & \multicolumn{4}{c}{\textbf{Number of Images}} & \multicolumn{2}{c}{\textbf{Original Image}} & \textbf{Training} \\ \cmidrule(r){3-6} \cmidrule(r){7-8} \cmidrule(r){9-9}
 &  & \textbf{Training} & \textbf{Testing} & \textbf{Total} & \textbf{Augmented} & \textbf{Resolution} & \textbf{FOV} & \textbf{Details} \\ \midrule
\multirow{3}{*}{Vessels} & DRIVE \cite{qureshi2013manually} & 20 & 20 & 40 & 720 & 565 $\times$ 584 & 35 &  \multirow{3}{*}{Image Level}  \\
 & CHASE\_DB \cite{7530915} & 28 & - & 28 & 720 & 1024 $\times$ 1024 & 45  \\
 & STARE \cite{STAREDataset} & 20 & - & 20 & 720 & 700 $\times$ 605 & 45  \\ 
\multirow{1}{*}{A/V}  & RITE\cite{qureshi2013manually} & 20 & 20 & 40 & 720 & 565 $\times$ 584 & 35    \\ \bottomrule

\end{tabular}
}
\label{tab:datasets}
\end{table*}


\subsection{Implementation Details}
Our model is implemented in TensorFlow and Keras and an NVIDIA Tesla P100 GPU with 32 GB RAM is used to perform experiments with batch size 8. The model is also evaluated on multi-class segmentation, which performs retinal artery and vein segmentation from retinal images on the RITE dataset \cite{hu2013automated}. We employed 80\% images for model training and 20\% validation from each dataset.

\subsection{Performance Metrics}\label{PM}
Standard evaluation metrics, including accuracy, dice, jaccard, specificity, sensitivity, and area under the curve (AUC), are used to assess the model performance.
Below are the specified evaluation metrics:
\begin{equation}
Accuracy(acc) = \frac {TP + TN}{TP + TN + FP + FN}
\end{equation}
\vspace{-0.1cm}
\begin{equation}
dice = \frac {TP + TP}{TP + TP + FP + FN}
\end{equation}
\vspace{-0.2cm}
\begin{equation}
Jaccard(J) = \frac {TP}{FN + FP + TP}
\end{equation}
\vspace{-0.2cm}
\begin{equation}
Sen = \frac {TP}{FN + TP}
\end{equation}
\vspace{-0.1cm}
\begin{equation}
Sp = \frac {TN}{FP + TN}
\end{equation}
\begin{equation}
AUC = \frac{1}{2 \times TP \times TN} \sum_{i=1}^{TP} \sum_{j=1}^{TN} \left( 1 + \frac{FN}{FP} \right)
\end{equation}
where, false positive, true positive, false negative, and true negative are represented by \(FP, TP, FN, TN\), respectively. Moreover, $Sp$ and $Sn$ represent specificity and sensitivity, respectively.

\subsection{Performance Comparisons of the Blood Vessels Segmentation}\label{subsec:comparisionsBV}

The quantitative assessment of our suggested model is given in Table \ref{tab:Vessels} along with other existing models. The table shows that LVS-Net outperforms existing techniques in several performance criteria, including sensitivity, specificity, accuracy, and dice score, while retaining the advantage of being lightweight. LVS-Net specifically achieved performance of 96.64\%, 86.44\%, 76.16\%, 83.91\%, and 98.51\% for accuracy, dice, jaccard, sensitivity, and specificity on the DRIVE dataset. Similarly, on the STARE dataset, it achieved 97.59\%, 84.78\%, 71.56\%, 85.19\%, and 98.61\% for accuracy, dice, jaccard, sensitivity, and specificity, respectively. In the CHASE\_DB dataset, it achieved scores of 96. 44\%, 84. 78\%, 73. 65\%, 83. 29\%, and 98. 44\% for accuracy, dice, Jaccard, sensitivity, and specificity, respectively. These results show the superior performance of LVS-Net over other state-of-the-art models. \\

\begin{table*}[!t]
  \centering
 \caption{Performance comparison of LVS-Net with recent methods on the DRIVE, STARE, and CHASE\_DB datasets. Various performance measures are highlighted, including accuracy, Dice coefficient, Jaccard index, sensitivity, and specificity, with the best results emphasized in bold for clarity.}
   \adjustbox {max width=\textwidth}
   {
\begin{tabular}{lcccccccccccccccccc}
    \toprule
    \multirow{3}[6]{*}{\textbf{Method}} & \multirow{3}[6]{*}{\textbf{Param (M)}} & \multicolumn{17}{c}{\textbf{Performance Measures in (\%)}} \\
\cmidrule{3-19}          &       & \multicolumn{5}{c}{\textbf{DRIVE}} &       & \multicolumn{5}{c}{\textbf{STARE}} &       & \multicolumn{5}{c}{\textbf{CHASE\_DB}} \\
\cmidrule{3-7}\cmidrule{9-13}\cmidrule{15-19}          &       & \textbf{Acc.} & \textbf{Dice}  & \textbf{J}&\textbf{Sn} & \textbf{Sp} &       & \textbf{Acc.} & \textbf{Dice} & \textbf{J} & \textbf{Sn} & \textbf{Sp} &       & \textbf{Acc.} & \textbf{Dice} & \textbf{J} & \textbf{Sn} & \textbf{Sp} \\
    \midrule
    
    U-Net \cite{ronneberger2015u}  & 7.76 & 96.78 &81.41  &68.64&80.57  &98.33  &       &  97.30& 81.18 &68.56&  70.50&  98.84&       & 97.43 &78.98  &65.26&76.50  & \textbf{98.84} \\
    G-Net Light \cite{iqbal2022g}& 0.39 & 96.86 & 82.02 & 69.09& 81.92 & 98.29 &       & 97.30 & 82.78 &69.64& 81.70 & 98.53 &       & 97.26 & 80.48 &67.76 &82.10 & 98.38 \\
    Att.Unet\cite{oktay2018attention}  & 9.25 &  96.62& 80.39 &67.21& 79.06 & 98.31 &       & 97.30 &81.06  &68.39&  78.04& \textbf{98.87} &       & 97.30 &79.64  &66.17& \textbf{84.84} &98.31  \\
    MultiResNet \cite{ibtehaz2020multiresunet}  & 7.20 & 95.64 & 82.32 &69.26& 79.46 & 97.89 &       & 96.33 & 82.44 &68.27& 77.09 & 98.48 &       & 96.42 & 80.12 &67.09& 80.10 & 98.04 \\
    BCD-UNet \cite{azad2019bi} & 20.65 & 95.75 & 82.49 & 69.33&79.84 & 98.03 &       & 96.34 & 82.30 &68.14& 78.92 & 98.16 &       & 96.18 & 79.32 &67.42& 77.35 & 98.01 \\
    SegNet \cite{badrinarayanan2017segnet} & 1.42 & 95.81 & 83.02 & 70.23&80.18 & 98.26 &       & 96.72 & 83.41 & 70.71&80.12 & 98.65 &       & 96.78 & 81.96 &68.56& 81.38 & 98.24 \\
    U-Net++\cite{zhou2018unet++} & 9.04 & 94.61 & 80.60 &68.27 &78.40 & 98.00 &       & 96.41 & 81.40 & 69.02&79.02 & 98.36 &       & 96.64 & 83.49 &66.88& 82.83 & 98.21 \\
    FR\_UNet\cite{liu2022full}   & 5.72 &  \textbf{97.05}&  83.16&71.20&  83.56&98.37  &       & \textbf{97.52} & 83.30 &72.46&  83.27& 98.69 &       & \textbf{97.48} &  81.51&68.82& 87.98 &  98.14\\
   RetinaLiteNet \cite{mehmood2024retinalitenet} & \textbf{0.066} & 95.07 & 80.60 &70.50&  78.40& 98.00 &       &  95.25& 80.87 & 71.03&  78.32& 98.04 &       & 95.57 &  78.53& 67.32&  73.12& 97.92 \\
    \midrule
    \textbf{LVS-Net} & 0.71 & 96.64 & \textbf{86.44} & \textbf{76.16} &\textbf{83.91} & \textbf{98.51} &       & 97.39 & \textbf{87.88} & \textbf{78.40} &\textbf{87.40} & 98.60 &       & 96.44 & \textbf{84.78} &\textbf{73.65}& 83.43 & 98.19 \\
    \bottomrule
    \end{tabular}%

    }
  \label{tab:Vessels}%
\end{table*}%

\begin{figure*}
    \centering
     \includegraphics[width=1\textwidth]{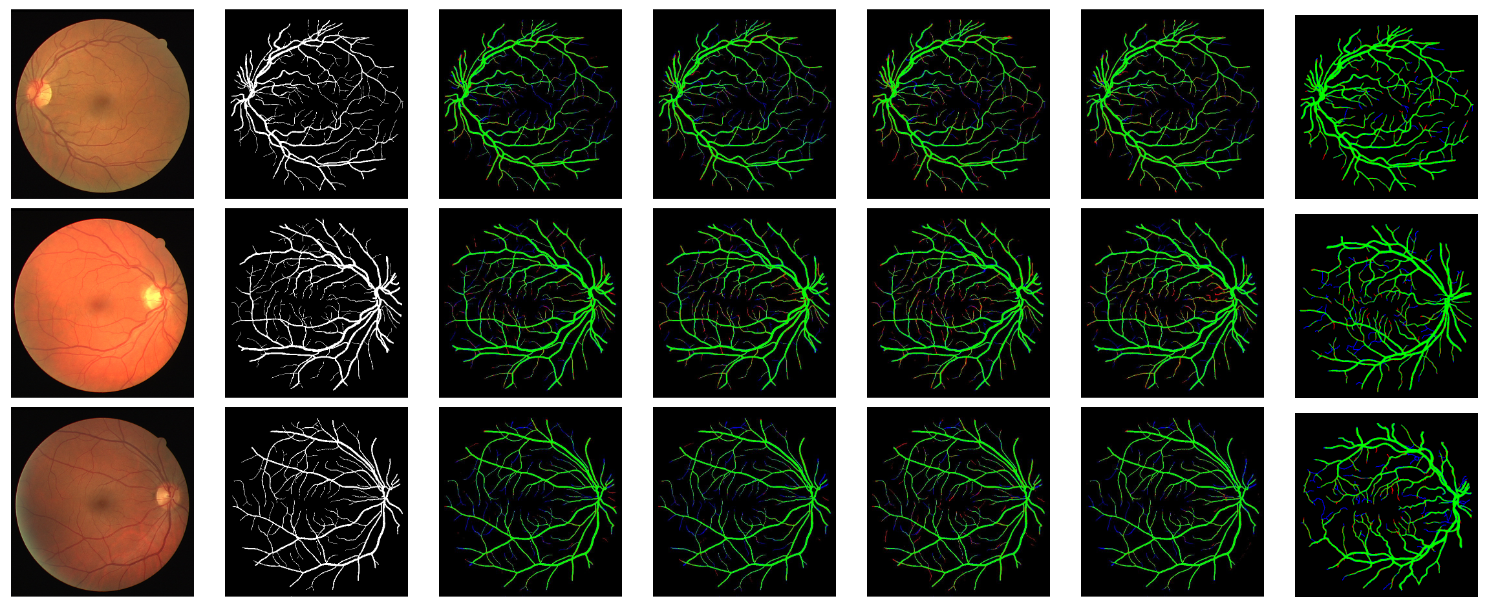}
    \caption{Segmentation outcomes of selected test images from the DRIVE dataset. Arranged in a left-to-right sequence are the input images, specifically images 1, 2, and 19 from the dataset. Following the input images are the ground truth and the outputs of the G-Net Light, MultiResNet, SegNet, U-Net++, and LVS-Net models. True positives are shown with green color, red pixels represent false positive detections, whereas blue pixels indicate false negative detections.}
    \label{visualDRIVE}
\end{figure*}

A clear assessment of the performance of the prediction algorithm is presented in Fig.~\ref{visualDRIVE} by marking true positives with green, false positives with red, and false negatives with blue. Visual inspection of the results in the DRIVE dataset reveals that the proposed LVS-Net yielded fewer false positives on thin vessels compared to the recent methods. Furthermore, the U-Net variants struggled with the boundaries (as depicted in image 4). SegNet generated false tiny vessels in most images, and G-Net Light tended to skip vessel information, which was robustly captured by LVS-Net while concealing inaccurate vessel information.

\begin{figure*}
    \centering
     \includegraphics[width=1\textwidth]{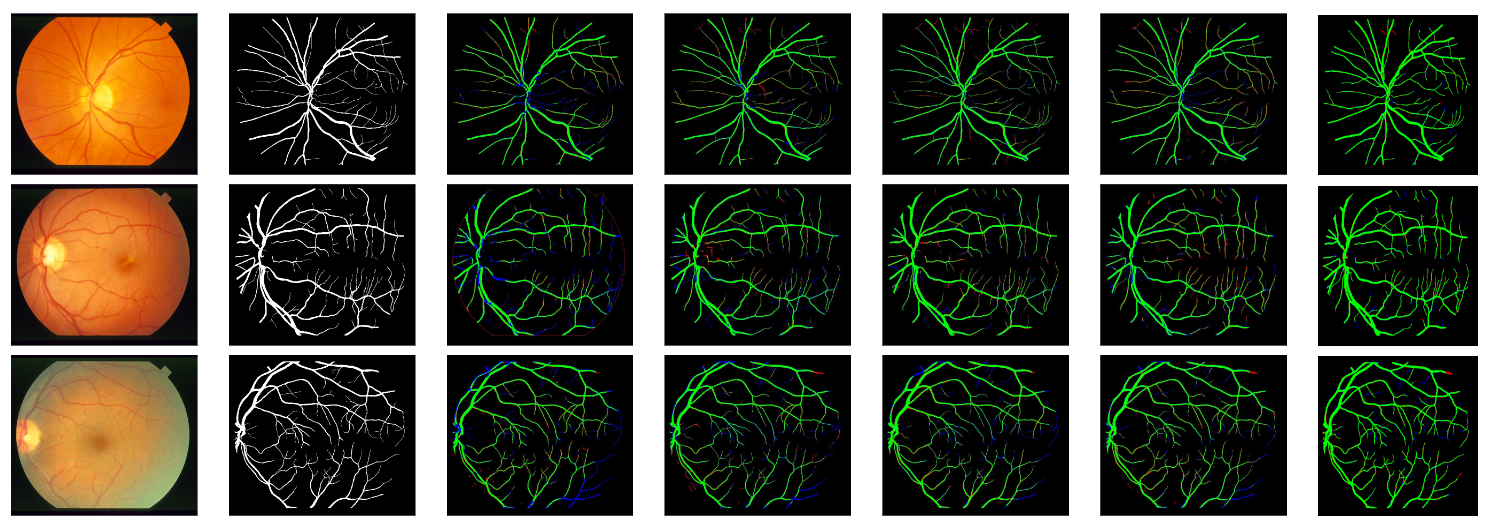}
    \caption{Segmentation outcomes of selected test images from the STARE dataset. The images displayed in the following order are: input images (specifically images 2, 3, and 5 from the dataset). Following the input images are the ground truth and the outputs of the G-Net Light, MultiResNet, SegNet, U-Net++, and LVS-Net models. True positives are shown with green color, red pixels represent false positive detections, whereas blue pixels indicate false negative detections.}
    \label{visualSTARE}
\end{figure*}

\begin{figure*}
    \centering
     \includegraphics[width=1\textwidth]{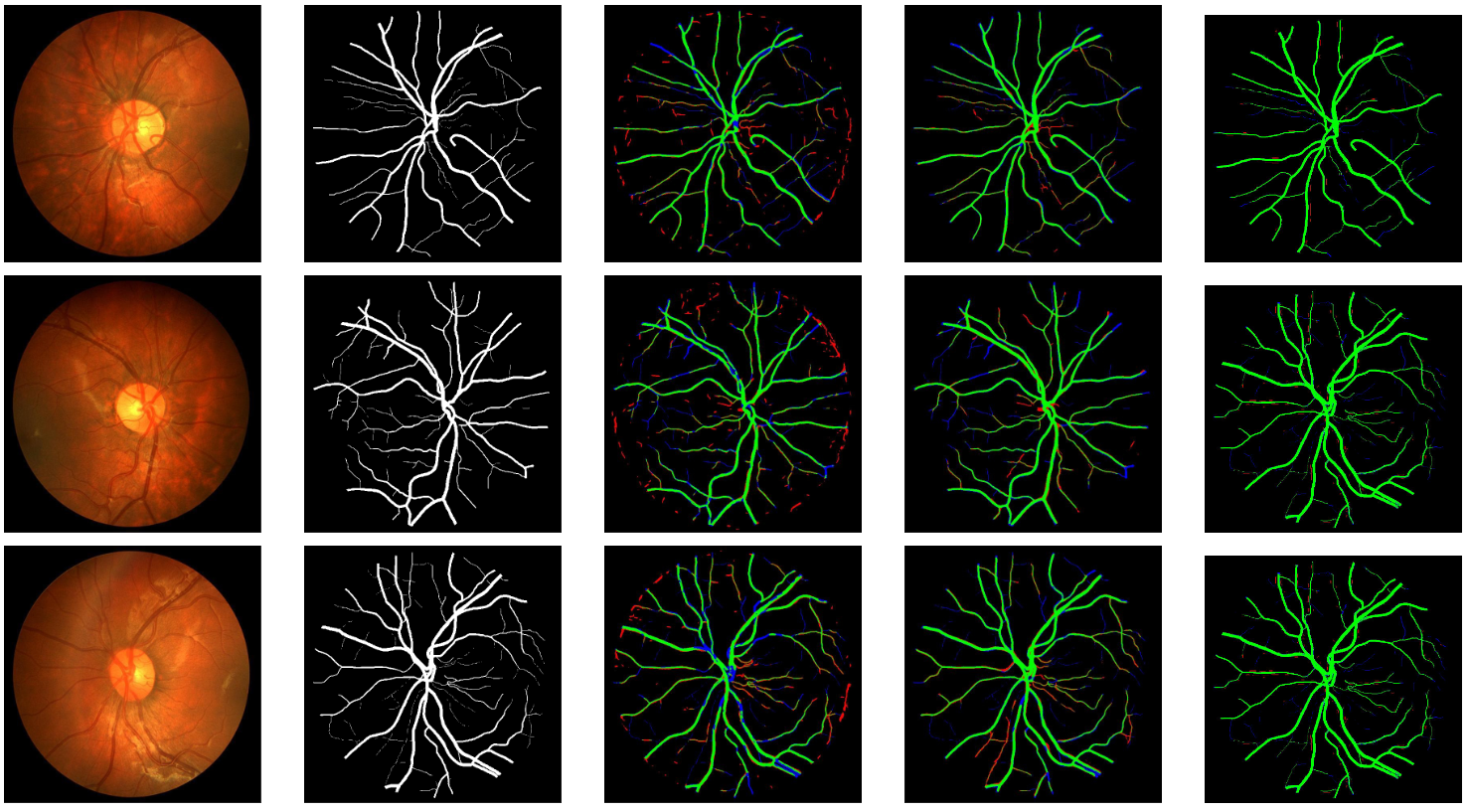}
    \caption{Segmentation outcomes of selected test images from the CHASE\_DB dataset \cite{CHASEDataset}. Arranged horizontally, the images displayed are as follows: the input images (specifically, images 1, 2, and 3 from the dataset). Following the input images are the ground truth and the outputs of the G-Net Light, SegNet, and LVS-Net models. True positives are shown with green color, red pixels represent false positive detections, whereas blue pixels indicate false negative detections.}
    \label{visualCHASE}
\end{figure*}

Similarly, inspection of the results in the STARE dataset Fig.~\ref{visualSTARE}, notably images 2 and 3) shows that the alternative methods produced more false positives, particularly across retinal boundaries, optic nerves, and tiny vessels. On the other hand, the proposed LVS-Net showed much more robustness to these artifacts in these images. Similar results can be observed from the results on the CHASE\_DB dataset in Fig.~\ref{visualCHASE}.\\

\begin{figure*}[!t]
	\centering
 \includegraphics[width = \textwidth]{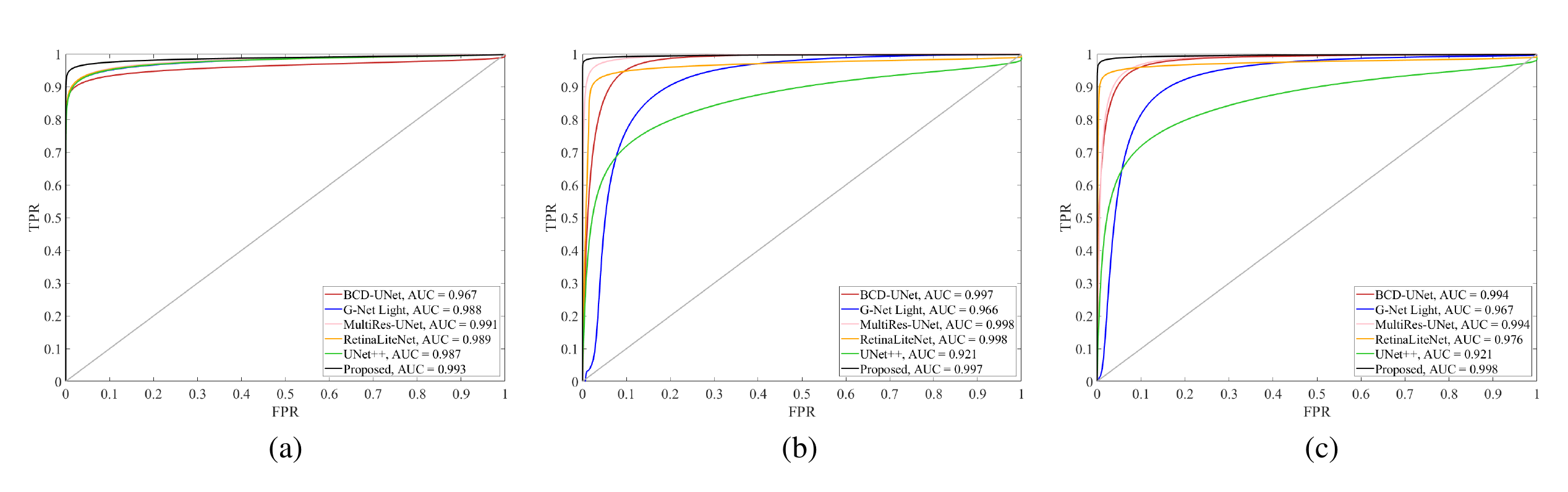}
	\caption{Comparative Analysis of Receiver Operating Characteristic (ROC) Curves for Different Models Evaluated on Three Retinal Datasets: (a) DRIVE, (b) STARE, and (c) CHASE\_DB with AUC Scores for BCD-UNet, G-Net Light, MultiRes-Unet, RetinaLightNet, UNet++ and LVS-Net.}
	\label{ROC_Curves}%
\end{figure*}%

In Fig. \ref{ROC_Curves}, the Receiver Operating Characteristic (ROC) curves are compared for three datasets, that is, DRIVE, STARE, and CHASE\_DB, for various models together with our proposed model. Each figure displays the True Positive Rate (TPR) against the False Positive Rate (FPR). Outperforming the other models, the proposed model consistently obtains the highest AUC values across all datasets. In particular, the suggested model outperforms the existing models in terms of AUC with values of 0.993, 0.997, and 0.998 for DRIVE, STARE, and CHASE\_DB, respectively. 

\subsection{Performance Comparisons of the AV Segmentation}\label{subsec:comparisons}

This section evaluates the generalization of our model by performing retinal artery vein segmentation using the RITE dataset. The findings demonstrate that our model obtained a higher dice score compared to other models on the RITE dataset. Consequently, our model demonstrates performance comparable to that of existing models while maintaining the advantage of being lightweight.\\

\begin{table*}[h!]
  \centering
 \caption{Comprehensive Performance Matrix Comparison of LVS-Net and other existing Models (U-Net++, Att-Unet, BCD-Unet, FR\_Unet, RetinalNet) on the RITE Dataset: Analysis of Accuracy, Dice Score, Sensitivity, and Specificity Across Arteries, Veins, and Overall Metrics.}
  \resizebox{\textwidth}{!}{%
    \begin{tabular}{llcccccccccccccc}
    \toprule
    \multirow{3}[6]{*}{\textbf{Dataset}} & \multirow{3}[6]{*}{\textbf{Model}} & \multicolumn{14}{c}{\textbf{Performance Measures (\%)}} \\
\cmidrule{3-16}          &       & \multicolumn{4}{c}{\textbf{Arteries}} &       & \multicolumn{4}{c}{\textbf{Veins}} &       & \multicolumn{4}{c}{\textbf{Average}} \\
\cmidrule{3-6}\cmidrule{8-11}\cmidrule{13-16}          &       & \textbf{Acc.} & \textbf{Dice} & \textbf{Sn} & \textbf{Sp} &       & \textbf{Acc.} & \textbf{Dice} & \textbf{Sn} & \textbf{Sp} &       & \textbf{Acc.} & \textbf{Dice} & \textbf{Sn} & \textbf{Sp} \\
    \midrule
    \multirow{6}[4]{*}{\textbf{RITE}} & UNet++ \cite{zhou2018unet++} & 97.07 & 67.47 & 70.06 & 98.30 &       & 97.56 & 51.76 & 61.24 & \textbf{98.84} &       & 95.28 & 73.41 & 70.32 & 97.96 \\
          & Att.Unet \cite{oktay2018attention} & 97.10 & 73.54 & 72.06 & 98.78 &       & 97.46 & 63.66 & 68.83 & 98.08 &       & 95.72 & 78.29 & 70.99 & 97.89 \\
          & BCD-Unet \cite{azad2019bi}  & 97.11 & 71.00 & 69.98 & 98.76 &       & 97.67 & 66.77 & 68.25 & 97.68 &       & 94.87 & 76.24 & 69.29 & 97.55 \\
          & FR\_Unet \cite{liu2022full} & 97.12 & 66.69 & 73.49 & 98.76 &       & 97.15 & 61.77 & 66.65 & 98.43 &       & 95.13 & 71.72 & 70.32 & 97.82 \\
          & RetinalNet \cite{mehmood2024retinalitenet} & 97.05 & 74.55 & 71.32 & 98.70 &       & 97.51 & 70.11 & 70.57 & 98.41 &       & 94.50 & 70.12 & 71.82 & 97.91 \\
\cmidrule{2-16}          & \textbf{LVS-Net} & \textbf{97.13} & \textbf{75.46} & \textbf{73.52} & \textbf{99.95} &       & \textbf{99.75} & \textbf{71.18} & \textbf{75.48} & \textbf{98.44} &       & 98.44 & \textbf{81.34} & \textbf{81.83} & \textbf{99.21} \\
    \bottomrule
    \end{tabular}%
    }

\label{tab:combined}
\end{table*}%

Table \ref{tab:combined} shows the performance of our model, and it is a comparison with existing state-of-the-art models tested on the RITE dataset. The table presents a comprehensive performance comparison of several retinal feature segmentation models, including UNet++, Att.Unet, BCD Unet, FR-UNet, RetinaLiteNet, and LVS-Net across the above mentioned datasets. The performance metrics evaluated are accuracy, dice coefficient, sensitivity, and specificity.\\

LVS-Net demonstrates comparable performance across all datasets with a small parameter count. In the RITE dataset, LVS-Net achieves an average value between arteries, veins, and the background of 97.54\%, 81.34\%, 81.30\%, and 92.83\% of accuracy, dice, sensitivity, and specificity, respectively. For better insight, we have reported the individual values of performance matrices for arteries and veins. These values are the performance measure between the arteries and the background for arteries and veins and the background for the veins. These results indicate its effectiveness in segmenting both arteries and veins, with significant improvements over other existing models.\\


LVS-Net consistently performs better than the other models on all datasets, demonstrating its reliability and accuracy in retinal arteries veins segmentation tasks. The equivalent values of performance matrices are attained by LVS-Net, demonstrated by the average performance metrics. These findings suggest that LVS-Net is an important tool in medical imaging analysis, particularly well-suited for applications demanding accurate feature segmentation. \\

Our model consistently achieves higher accuracy and dice scores due to its more complex architecture, which effectively captures the vessels and identifies the arteries and veins in the retinal images. From the table, we can see that the lightweight characteristics of RetinaLiteNet reduced its ability to handle these complexities, resulting in performance degradation in multi-class problems.\\

The performance of traditional models UNet variants serves as a solid baseline, but the advances in models such as LVS-Net provide substantial improvements in vessel segmentation. The LVS-Net performance metrics are the most effective model for this task and offer a robust foundation and development in this domain for future research.

\subsection{Ablation Study on DRIVE dataset}

The ablation study, presented in Table \ref{tab:Ablation}, performed on the DRIVE dataset, provides insight into the effects of various components on model performance. Evaluate the addition of various components to the lightweight U-Net (LU) baseline model to determine how those components affect the final model. When multiscale layers (MLU) are integrated, overall performance metrics are slightly improved, starting with LU. CBAM is added to the skip connections of MLU, substantially improving specificity and accuracy, highlighting the importance of attention mechanisms. Performance is improved by implementing $\texttt{SFRB}$ in skip connections and the bottleneck. The dice, jaccard, and sensitivity gradually increased. The ultimate model achieves the best performance across all metrics by implementing $\texttt{SFRB}$ in skip connections and combining the components $\texttt{SFRB}(\mathcal{G})$ and $\texttt{FMAM}(\mathcal{F})$ on bottlenecks. These modifications indicate that the cumulative effect of these improvements significantly improves segmentation abilities. \\

\begin{table*}[!t]
  \centering
  \caption{Performance Comparison of LVS-Net with Various Retinal Feature Segmentation Models on RITE Dataset Across Different Metrics (Accuracy, Dice Score, Sensitivity, and Specificity) for Arteries, Veins, and Combined Averages.}
   \resizebox{\textwidth}{!}{%
    \begin{tabular}{lccccc}
    \toprule
    \multirow{2}[4]{*}{\textbf{Method}} & \multicolumn{5}{c}{\textbf{Performance Measures (\%)}} \\
\cmidrule{2-6}          & \textbf{Dice} & \textbf{J} & \textbf{ Acc} & \textbf{Sen} & \textbf{Sp} \\
    \midrule
    Lightweight U-Net (LU) & 82.06 & 66.29 & 96.09 & 82.20 & 98.04 \\
    Multiscale LU (MLU) ($3\times3$, $1\times1$) & 83.49 & 68.57 & 96.52 & 83.63 & 98.47 \\
    MLU + CBAM in Skip Connections & 80.86 & 67.53 & 95.80 & 78.42 & 97.41 \\
    MLU + ($\mathcal{G}$) in Skip Connections (($\mathcal{G}$)-Skip) & 81.99 & 68.98 & 96.82 & 82.07 & 98.25 \\
    MLU + ($\mathcal{G}$)-Skip + ($\mathcal{G}$) in Bottleneck (($\mathcal{G}$)-Bottleneck) & 82.24 & 69.90 & \textbf{96.89} & 82.49 & 98.28 \\
    MLU + ($\mathcal{F}$) in Skip Connections (($\mathcal{F}$)-Skip) & 82.25 & 70.23 & 95.96 & 81.48 & 97.42 \\
    MLU + ($\mathcal{F}$) in Bottleneck (($\mathcal{F}$)-Bottleneck) & 83.81 & 73.11 & 96.17 & 83.18 & 97.77 \\
    \midrule
    \textbf{MLU + ($\mathcal{G}$)-Skip + ($\mathcal{G}$)-Bottleneck + ($\mathcal{F}$)-Bottleneck} & \textbf{86.44} & \textbf{76.16} & 96.64 & \textbf{83.91} & \textbf{98.51} \\
    \bottomrule
    \end{tabular}%
    }
  \label{tab:Ablation}%
\end{table*}%

\begin{figure*}
    \includegraphics[width = \textwidth]{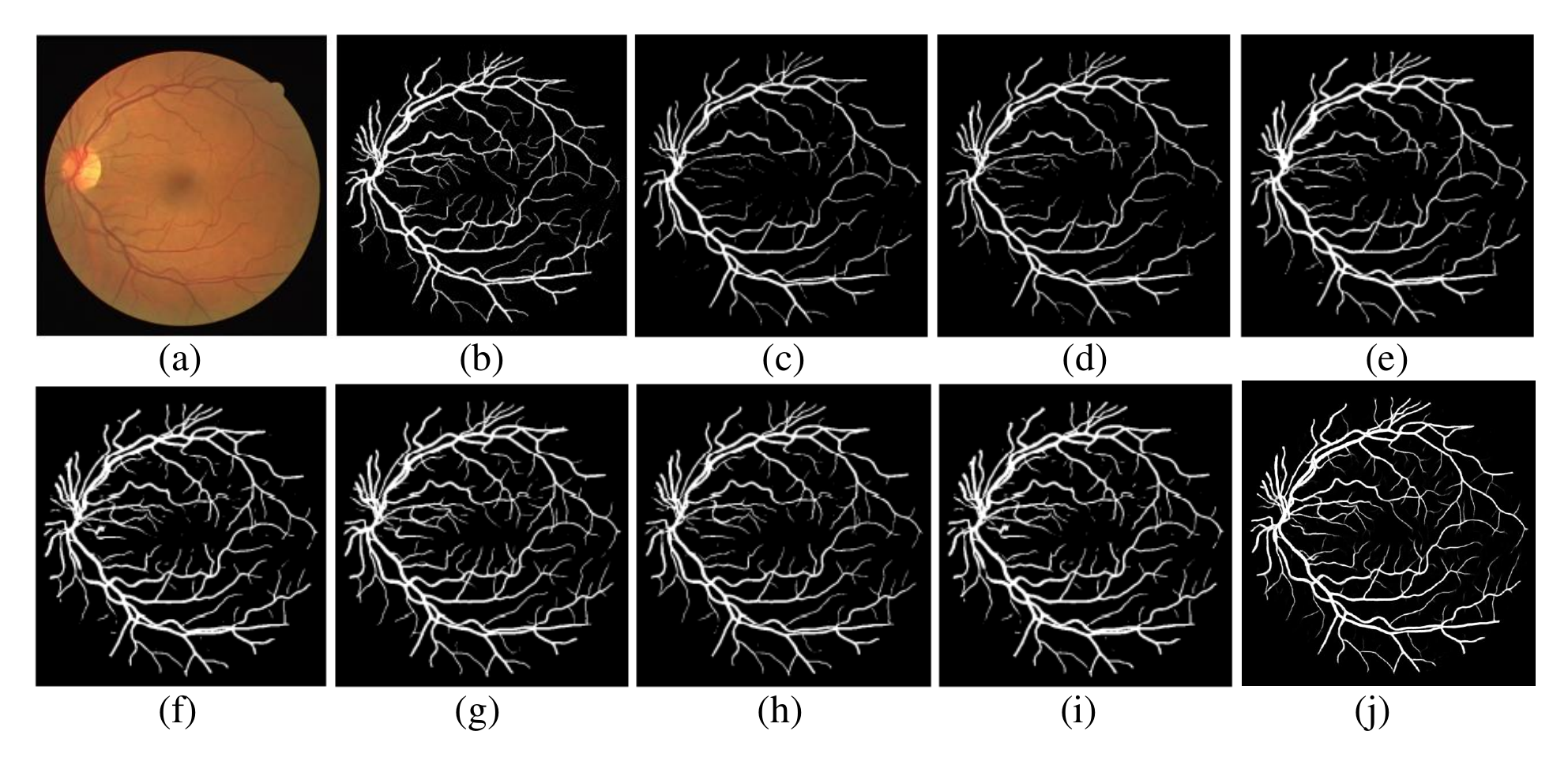}
    \caption{Illustration of the visual results obtained by employing different components of the proposed architecture: (a) Input RGB image, (b) Corresponding ground-truth images, (c) Lightweight U-Net (LU), (d) Multiscale LU (MLU) ($3\times3$, $1\times1$), (e) MLU + CBAM in Skip Connections, (f) MLU + ($\mathcal{G}$) in Skip Connections (($\mathcal{G}$)-Skip), (g) MLU + ($\mathcal{G}$)-Skip + ($\mathcal{G}$) in Bottleneck (($\mathcal{G}$)-Bottleneck), (h) MLU + ($\mathcal{F}$) in Skip Connections (($\mathcal{F}$)-Skip), (i) MLU + ($\mathcal{F}$) in Bottleneck (($\mathcal{F}$)-Bottleneck), and (j) MLU + ($\mathcal{G}$)-Skip + ($\mathcal{G}$)-Bottleneck + ($\mathcal{F}$)-Bottleneck}
    \label{fig:ablation}
\end{figure*}
    \label{fig:ablation}
The images in Fig. \ref{fig:ablation} visually demonstrate the impact of various components added to the proposed model. We can see that images (c), (d), and (e) depict results from Lightweight U-Net (LU), Multiscale LU (MLU), and MLU with CBAM in Skip Connections, respectively, starting with (a) as the input RGB image and (b) as the ground truth. There are gaps in the vessel continuity and insufficient segmentation of the tiny vessels in these early models. Segmentation improves as we move on to images (f) and (g), which employ $\texttt{SFRB}$ to skip connections and bottlenecks. Despite this, there is still some dilatation visible in the vessels. Using $\texttt{FMAM}$ in the skip connections and bottleneck of images (h) and (i), the features of the vessel are further refined, capturing the delicate structures with less dilatation. The final image (j) obtains the optimal segmentation result, which depicts the entire model with the integrated components of MLU, $\texttt{SFRB}(\mathcal{G})$, and $\texttt{FMAM}(\mathcal{F})$. It is identical to the ground truth and effectively captures the small vessels without dilation, which represents the efficiency of the proposed model. 

\section{Conclusion}\label{Conclusion}
In this paper, we introduce a lightweight encoder-decoder model based on the segmentation of retinal blood vessels. Our model comprises an encoder and decoder, embedding multi-scale convolutional layers with the combination of focal modulation attention and the spatial feature refinement blocks at the bottleneck employed for the feature extraction. Furthermore, both the decoder and the skip connections use the spatial feature refinement block, helping to highlight and enhance key features. It is beneficial for image segmentation tasks where precise segmentation and localization of objects are important. Promising findings have been obtained from an extensive analysis of the model on public datasets, that is, DRIVE, CHASE\_DB and STARE datasets, generating dice scores of 86. 44\%, 82. 10\% and 87. 88\%, respectively. Our model satisfies its lightweight requirements, 2.74 MB of memory, 0.71 million parameters, and 29.60 GFLOPs. These findings demonstrate that significant medical image analysis persists and is feasible even with limited hardware resources.\\


\end{document}